\documentclass[useAMS,usenatbib]{mn2e}
\usepackage{graphicx}
\usepackage{rotating}

\newcommand{\revised}[1]{{#1}}

\title[HFQPO from GRS 1915+105]{High-Frequency Quasi-Periodic Oscillations from GRS 1915+105}
\author[T. M. Belloni \& D. Altamirano]{T.M. Belloni$^{1}$\thanks{E-mail:
tomaso.belloni@brera.inaf.it}, D. Altamirano$^{2}$\\
$^{1}$INAF - Osservatorio Astronomico di Brera, Via E. Bianchi 46, I-23807, Merate, Italy\\
$^{2}$Astronomical Institute Anton Pannekoek, University of Amsterdam, Postbus 94249, 1090 GE Amsterdam, the Netherlands}

\begin{document}

\date{Accepted 2013 March 18.  Received 2013 March 18; in original form 2012 October 2}

\pagerange{\pageref{firstpage}--\pageref{lastpage}} \pubyear{2012}

\maketitle

\label{firstpage}

\begin{abstract}
We report the results of a systematic timing analysis of all archival Rossi X-Ray Timing Explorer (RXTE) observations of 
the bright black-hole binary GRS 1915+105 in order to detect high-frequency quasi-periodic oscillations (HFQPO). We produced power-density spectra in two energy bands and limited the analysis to the frequency range 30-1000 Hz.
We found \revised{51} peaks with a single trial significance larger than 3$\sigma$. As all but three have centroid frequencies that are distributed between 63 and 71 Hz, we consider most of them significant regardless of the number of trials involved. The average centroid frequency and FWHM are 67.3$\pm$2.0 Hz and 4.4$\pm$2.4 Hz respectively. Their fractional rms varies between 0.4\% and 2\% (total band detections) and between 0.5\% and 3\% (hard ban detections).
As GRS 1915+105 shows large variability on time scales longer than 1s, we analysed the data in 16s intervals and found that the detections are limited to a specific region in the colour-colour diagram, corresponding to state B of the source, when the energy spectrum is dominated by a bright accretion disk component. However, the rms spectrum of the HFQPO is very hard and does not show a flattening up to 40 keV, where the fractional rms reaches 11\%.
We discuss our findings in terms of current proposed models and compare them with the results on other black-hole binaries and neutron-star binaries.

\end{abstract}


\section{Introduction}

The large wealth of X-ray observations of black-hole transients (BHTs) accumulated by the Rossi X-Ray Timing Explorer (RXTE) mission has allowed us to obtain a much more precise view of the properties of accretion onto stellar-mass black holes (see e.g. Belloni 2010; Fender 2010; Belloni et al. 2012) compared to the sparse observations of a few objects which were available before.
In addition to detailed information on low-frequency (0.1-30 Hz) Quasi-Periodic Oscillations (QPO, see e.g. Motta et al. 2011), RXTE led to the discovery of higher-frequency features, which sample the range expected from signals associated to Keplerian motion in the innermost regions of an accretion disk around a black hole.
These QPOs at higher frequencies (30-450 Hz) are known as High-Frequency QPOs (HFQPOs) and are mostly
weak and elusive signals. Although the existing database of RXTE observations is very large, only a small number of detections is available, associated only to a few sources: GRS 1915+105 (Morgan et al. 1997; Strohmayer 2001a; Belloni et al. 2001; Remillard et al. 2002b; Belloni et al. 2006), GRO J1655-40 (Remillard et al. 1999; Strohmayer 2001b), XTE J1550-564 (Homan et al. 2001; Miller et al. 2001; Remillard et al. 2002a), H1743-322 (Homan et al. 2005; Remillard et al. 2006), XTE J1650-500 (Homan et al. 2003), 4U 1630-47 (Klein-Wolt, Homan \& van der Klis 2004), XTE J1859+226 (Cui et al. 2000) and IGR J17091-3624 (Altamirano \& Belloni 2012). Some of these detections appear in pairs.
Recently, Belloni, Sanna \& M\'endez (2012) examined systematically the full archive (with the exception of two sources, see below) and found only eleven statistically-significant detections from two sources, after taking into account the number of trials. All detections appear when the sources are in an intermediate state and none of them together with a type-C LFQPO (see below). For all detections, there is an indication that there are preferred frequencies, but the small number of detections prevents a more solid assessment.

Low-Frequency QPOs (LFQPOs) are much more common in BHTs (see Motta et al. 2011 and references therein). They have been classified into three separate types (called A, B and C). Type-C QPOs are the most commonly observed. They vary in frequency over a rather large range and have been interpreted as related to the frame-dragging precession time scale (see e.g. Stella, Vietri \& Morsink 1999; Ingram, Done \& Fragile 2009). Type-B QPOs are restricted over  a much small range of frequencies, while type-A QPOs are much fainter signals and less information is available.

Two sources were not considered by Belloni, Sanna \& M\'endez (2012). IGR J17091-3624 and GRS 1915+105, as their peculiar behaviour requires a special analysis.  IGR J17091-3624 was analysed in detail by Altamirano \& Belloni (2012). Here we concentrate on GRS 1915+105 and present a systematic high-frequency timing analysis of the full set of existing RXTE observations of BHTs. 

\section{GRS 1915+105}

GRS 1915+105, discovered in 1992 as a bright transient which has remained active since, is a bright black hole binary which displays a very peculiar behaviour in the form of complex structured variability (see Belloni et al. 2000; Fender \& Belloni 2004 for a review). Its luminosity is estimated to be near Eddington and it is the most important system to study the accretion-ejection connection.
Its variability was studied by Belloni et al. (2000), who classified its behaviour into 12 variability classes, extended later to 14 (Klein-Wolt et al. 2002; Hannikainen et al. 2005).
It was a unique source until in 2011 a new much fainter system was discovered, IGR J17091-3624, which displayed the very same type of peculiar variability (see Altamirano et al. 2012 and references therein).

As mentioned above, high-frequency oscillations were seen with RXTE from GRS 1915+105. Morgan et al. (1997) reported the discovery of a HFQPO around 65-67 Hz, which varied little between the very few observations where it was seen. Its total rms amplitude was low, around 1\%, with a hard spectrum (up to 6\% in the 15-25 keV band). Its quality factor $Q$, defined as the ratio between centroid frequency and FWHM was high, around 20.
Belloni et al. 2001 analysed in detail two of these observations and found that the spectrum of the 67 Hz QPO changed dramatically as a function of a much slower 0.067 Hz oscillation, and discovered a broader (Q$\sim$3) QPO at 27 Hz, also dependent on the slower oscillation. Around the same time, Strohmayer (2001a) discovered a 41 Hz QPO when analyzing new observations taken in 1997 where a 69 Hz QPO was detected. In this case, the 41 Hz and 69 Hz peaks were detected simultaneously. It was noticed that the ratio between these frequencies approached 5:3 (Kluzniak \& Abramowicz 2002).
Noticeably, the sequence 27:41:69 approaches the 2:3:5 ratios.
Remillard et al. (2002b) reported the discovery of two additional peaks at 164 and 328 Hz (consistent with being harmonics), from a selection made in hardness and intensity from a set of RXTE observations in 1997. Belloni et al. (2006) reanalysed the data and confirmed the $\sim$170 Hz , although with a rather low Q$\sim$2.

\section{Data and analysis}

We selected all RXTE observations of GRS 1915+105 available in the archive from the 
start to the end of the mission, concentrating on the data from the Proportional Counter Array (PCA) instrument. 
We analyzed a total of 1807 observations, for a total of 5181 ks exposure. Given the size of the available database, we followed a semi-automatic procedure similar to that followed by Belloni, Sanna \& M\' endez (2012):

\begin{enumerate}

\item We produced Power Density Spectra (PDS) from two energy bands: channels 0-79 and 14-79 (total and hard band respectively, corresponding to energies \revised{1.9-29.8 keV and 5.5-29.8 keV for the first observations where we had a detection (see below) and to energies 2.1-33.4 keV and 6.1-33.4 keV} at the end of the mission, due to gain changes in the detectors. For some observations, we could not use these channels bands due to incompatible data modes. In these cases, we chose the closest approximation to the desired bands. In a few cases, the data modes were such that we used the full channel range and/or the total and hard band were the same.
The choice of the hard band was made in order to include a sufficient number of photons and to allow the examination of more observations (see next bullet).
We consider a separate observation (or dataset) the data corresponding to a RXTE observation ID.

\item The large number of observations made it necessary to perform an automatic search in order to select candidates for HFQPOs. For each observation, the adopted procedure, applied both to the total and hard energy band, was the following:

\begin{itemize}

\item The dataset was subdivided into intervals of 16s duration, from each interval we produced a PDS and averaged all PDS. The time resolution for all spectra was 4096 points per second, corresponding to a Nyquist frequency of $\sim$2 kHz. The PDS were normalised according to Leahy et al. (1983) and rebinned in such a way that each frequency bin was larger than the previous one by $\sim$2\%.
Uncertainties in power were estimated following van der Klis (1988).

\item Since our interest was only in the high-frequency region, we limited our analysis to the frequency range 30-1000 Hz.
Extending the threshold to lower frequencies (where the 27 Hz features was detected) would have created problems due to residual source noise, which would complicate the automatic detection.

\item We fitted the PDS with a model consisting of the sum of a power law (to account for the Poissonian noise component) and a Lorentzian. The initial procedure was completely automatic. The slope of the power law was not fixed to 0  in order to fit any possible remaining tail of low-frequency features over the roughly flat Poisson level. We first fixed the Lorentzian centroid frequency to values between 30 and 1000 Hz in steps of 1 Hz, limiting the FWHM between 0.5 Hz and 1000 Hz. Then, we made a fit with a free centroid around the frequency corresponding to the minimum chi square from the previous procedure (regardless of the quality of the fit), obtaining the best fit for each observation. This procedure is aimed at detecting a single HFQPO in each PDS and is \revised{unable} to detect multiple peaks. A successful fit was obtained for all observations at this stage, although for many cases the QPO detection was not significant.

\item We then estimated the significance of the detection by dividing the normalisation of the Lorentzian (corresponding to the integral over positive frequencies) by its 1$\sigma$ error on the negative side. We retained only the \revised{109} detections
where this significance was larger than 3.

\item All the \revised{109} detections were examined visually, since in many cases the fits did not make sense. For example, in some cases the FWHM resulted in a very broad or very narrow feature (hitting the imposed limits), or the centroid frequency itself was also close to the search limits, or finally there was remaining source noise above 30 Hz, leading to a spurious QPO detection.
In addition, we discarded all fits where the QPO had a quality factor $Q=\nu_0 / FWHM$ significantly less than 2.

\item For the observations that survived the procedure, we repeated the fit manually and we estimated the significance of the fit. In a large number of cases, we found a peak at a consistent frequency both in the hard and total PDS. In these cases, we retained the parameters of the fit with the highest significance. No pairs of peaks at inconsistent frequencies between total and hard band were found. We were left with \revised{51} observations containing a significant peak. For all our final fits, the reduced chi square was close to unity. The log of observations is shown in Tab. \ref{tab:log} and the HFQPO parameters in Tab. \ref{tab:results}.

\end{itemize}

\item The final HFQPOs have a significance larger than 3$\sigma$, but for a single trial. To keep into account the number of trials in our procedure, we had to consider the number of observations analysed and the number of independent frequencies in the PDS.
As number of observations, we used 1807 if the QPO was detected in both bands, as the second detection did not add significance to the first. If the QPO was detected in only one of the two bands, we used 1807*2.
To estimate of the number of independent frequencies scanned for each observation where a QPO was detected, we divided the frequency range used for the analysis (970 Hz) by the FWHM of the QPO detected in that particular observation. 
Following this procedure, the number of trials results very large, of the order of $10^6$, which lowers considerably our sensitivity. Only 4 detections remain with \revised{a final detection probability} less than 1\%.
However, \revised{49 of our 51} detections fall in the frequency range \revised{58.9-71.3 Hz and 48 of them in the frequency range 63.5-71.3 Hz}, suggesting that most of them are real. Using that reduced frequency range (14 Hz) increases our $<$1\% observations to 14. 
We report all observations with single trial $>3\sigma$ as a posteriori the chance of having almost the totality of the QPOs within such a small range of frequencies is rather low.
\revised{Two detections are at higher frequency $>$100 Hz and have a high probability of being spurious after number of trials are taken into account. We include them in the following discussion not having strong statistical reasons to discard only these two values (we note that their frequency is consistent with being twice some values obtained in other observations, 67.05 Hz and 71.49 Hz). We checked whether these two observations also feature a lower-frequency HFQPO, but found none.}

\item Notice that our procedure can only detect one peak per PDS. In principle, two different features can be detected in the total and hard PDS, but this did not happen in any of our observations.

\end{enumerate}

\begin{table}
 \centering
 \caption{List of GRS 1915+105 observations with a HFQPO detection. Columns are: observation \#, observation ID, observation date (MJD), variability class, exposure time}
  \begin{tabular}{@{}lcccc@{}}
  \hline
   Obs. N &
   ObsID &
   MJD  &
   Class &
   Exp. (s)\\ 
 \hline

 1 & 10408-01-01-01 & 50179.236	&$\kappa$&	5600			\\ 
 2 & 10408-01-03-00 & 50190.578	&$\gamma$&	4768			\\
 3 & 10408-01-04-00 & 50193.439	&$\gamma$&	8768			\\
 4 &10408-01-05-00 & 50202.845	&$\gamma$&	9584			\\
 5 &10408-01-06-00 & 50208.584	&$\gamma$&	9856			\\
 6 &10408-01-07-00 & 50217.656	&$\gamma$&	9984			\\
 7 &10258-01-10-00 & 50351.668	&$\mu$&		5488			\\
 8 &20186-03-01-03 & 50628.615	&$\kappa$&	3504			\\
 9 &20186-03-01-04 & 50629.483	&$\kappa$&	10608		\\
10 &20402-01-38-00 & 50649.426	&$\gamma$&	7472			\\
11 &20402-01-39-00 & 50654.028	&$\gamma$&	6864			\\
12 &20402-01-39-02 & 50658.511	&$\gamma$&	2128			\\
13 &20402-01-41-02 & 50679.372	&$\delta$&	3312			\\
14 &20402-01-54-00 & 50763.209	&$\delta$&	10352		\\
15 &20402-01-55-00 & 50769.229	&$\delta$&	8768			\\
16 &20402-01-56-00 & 50774.224	&$\delta$&	9408			\\
17 &20402-01-57-01 & 50786.362	&$\mu$&		4672			\\
18 &20402-01-58-01 & 50788.366	&$\mu$&		4512			\\
19 &20402-01-60-00 & 50804.908	&$\delta$&	12368		\\
20 &30703-01-01-00 & 50810.027	&$\delta$&	5168			\\
21 &30402-01-01-00 & 50820.164	&$\delta$&	6880			\\
22 &40403-01-06-00 & 51158.274	&$\kappa$&	2144			\\
23 &40703-01-12-00 & 51288.149	&$\omega$&	9344			\\
24 &40403-01-07-00 & 51291.081	&$\omega$&	2816			\\
25 &40703-01-13-01 & 51299.134	&$\omega$&	6080			\\
26 &40703-01-26-00 & 51407.769	&$\omega$&	8320			\\
27 &40703-01-28-00 & 51418.781	&$\omega$&	1536			\\
28 &40703-01-28-02 & 51418.920	&$\omega$&	1856			\\
29 &40703-01-29-00 & 51426.769	&$\omega$&	1872			\\
30 &40703-01-29-01 & 51426.839	&$\omega$&	1504			\\
31 &40703-01-29-02 & 51426.909	&$\omega$&	1920			\\
32 &40703-01-31-00 & 51447.728	&$\delta$&	9984			\\
33 &50703-01-10-02 & 51681.213	&$\rho$&		864			\\
34 &50703-01-60-03 & 52052.234	&$\rho$&		1040			\\
35 &50703-01-61-01 & 52059.119	&$\rho$&		1648			\\
36 &60701-01-09-00 & 52192.641	&$\nu$&		6976			\\
37 &70702-01-55-00 & 53361.525	&$\delta$&	4832			\\
38 &80701-01-28-00 & 52933.627	&$\delta$&	1952			\\
39 &80701-01-28-01 & 52933.696	&$\delta$&	1616			\\
40 &80701-01-28-02 & 52933.764	&$\delta$&	1008			\\
41 &80127-04-02-00 & 52945.238	&$\delta$&	15104		\\
42 &80701-01-31-00 & 52955.570	&$\delta$&	11440		\\
43 &90105-08-02-00 & 53503.870	&$\delta$&	5104			\\
44 &90701-01-38-00 & 53347.564	&$\delta$&	1952			\\
45 &91701-01-12-00 & 53515.075	&$\delta$&	2112			\\
46 &91701-01-12-01 & 53515.140	&$\delta$&	2352			\\
47 &92092-03-01-00 & 53707.683	&$\delta$&	13088		\\
48 &93701-01-01-00 & 54285.950	&$\gamma$&	1648			\\
49 &93701-01-06-00 & 54320.822	&$\delta$&	3056			\\
50 &93411-01-01-00 & 54326.709	&$\delta$&	3424			\\
51 &95701-01-20-00 & 55335.140	&$\rho$&		1216			\\
 
\hline
\end{tabular}
\label{tab:log}
\end{table}

\begin{table*}
 \centering
\begin{minipage}{160mm}
 \caption{List of detected HFQPOs for GRS 1915+105. Columns are: observation number, bands in which the QPO was detected (Total/Hard: in boldface the most significant, from which the parameters were obtained), centroid frequency, FWHM and fractional rms of the QPO, chance probabilities for single trial (with number of sigma), full frequency range and restricted frequency range (see text), net source average observed count rate (in PCU2, with average number of active PCUs. Probabilities of 0 and 1 indicate values too close to those values to be displayed.}
  \begin{tabular}{@{}lcccccccc@{}}
  \hline
   Obs N &
   B&
   $\nu_0$ (Hz)&
   FWHM (Hz)&
   \%rms &
   P$_0$&
   P$_F$&
   P$_R$&
   Rate (N$_{PCU}$)     \\ 
 \hline

 1 &	{\bf T=H}	&	66.21	$_{-0.75	}^{+0.53	}$ &	3.11	$_{-2.11	}^{+1.60	}$ &	0.58	$_{-0.12	}^{+0.10	}$ &	1.35E-03	(3.00)&	1.00E+00	&	1.00E+00	&	4011	 (3.0)	\\
 2 &	{\bf T=H}	&	68.79	$_{-0.68	}^{+0.40	}$ &	2.86	$_{-2.50	}^{+2.11	}$ &	0.62	$_{-0.14	}^{+0.09	}$ &	1.59E-04	 (3.60)&	1.00E+00	&	7.55E-01	&	3818 (3.0)	\\
 3 &	{\bf T=H}	&	67.34	$_{-0.12	}^{+0.14	}$ &	3.47	$_{-0.50	}^{+0.44	}$ &	0.86	$_{-0.04	}^{+0.04	}$ &	0.00E+00	(12.10)&	0.00E+00	&	0.00E+00	&	3907	 (5.0)	\\
 4 &	T{\bf H}	&	66.31	$_{-0.29	}^{+0.31	}$ &	3.15	$_{-0.75	}^{+0.54	}$ &	0.95	$_{-0.09	}^{+0.07	}$ &	5.23E-12	(6.80)&	2.91E-06	&	4.20E-08	&	2502	 (5.0)	\\
 5 &	{\bf T}H	&	65.10	$_{-0.13	}^{+0.13	}$ &	4.08	$_{-0.38	}^{+0.29	}$ &	1.21	$_{-0.05	}^{+0.03	}$ &	0.00E+00	(17.60)&	0.00E+00	&	0.00E+00	&	3410	 (4.0)	\\
 6 & {\bf T}       &      66.98        $_{-0.50	}^{+0.61	}$ &	2.66	$_{-1.53	}^{+1.04	}$ & 	0.56	$_{-0.09	}^{+0.09	}$ & 1.22E-03 (3.03)   &   1.00E+00 &      1.00E+00 &      2951 (3.0)        \\
 7 &	T{\bf H}	&	67.71	$_{-0.59	}^{+0.58	}$ &	5.16	$_{-2.07	}^{+1.31	}$ &	0.41	$_{-0.08	}^{+0.07	}$ &	9.96E-08	 (5.2)0&	3.33E-02	&	4.89E-04	&	4305	 (5.0)	\\
 8 &	{\bf H}	&	63.54	$_{-1.42	}^{+1.38	}$ &	9.48	$_{-6.28	}^{+3.65	}$ &	1.17	$_{-0.29	}^{+0.20	}$ &	1.35E-03	(3.00)&	1.00E+00	&	9.99E-01	&	3187	 (5.0)	\\
 9 &	T{\bf H}	&	68.43	$_{-0.88	}^{+0.89	}$ &	10.63$_{-3.43	}^{+2.42	}$ &	1.06	$_{-0.14	}^{+0.11	}$ &	9.96E-08	(5.20)&	1.63E-02	&	2.37E-04	&	3258	 (5.0)	\\
10 &	{\bf T}H	&	67.40	$_{-0.60	}^{+0.59	}$ &	3.52	$_{-2.20	}^{+1.67	}$ &	0.42	$_{-0.08	}^{+0.06	}$ &	3.37E-04	(3.40)&	1.00E+00	&	9.11E-01	&	4053	 (5.0)	\\
11 &	T{\bf H}	&	68.92	$_{-0.21	}^{+0.21	}$ &	4.20	$_{-0.80	}^{+0.68	}$ &	1.13	$_{-0.07	}^{+0.06	}$ &	0.00E+00	(8.90)&	0.00E+00	&	0.00E+00	&	3882	 (4.7)	\\
12 &	T{\bf H}	&	68.92	$_{-0.41	}^{+0.39	}$ &	3.60	$_{-1.43	}^{+1.24	}$ &	0.98	$_{-0.13	}^{+0.11	}$ &	4.29E-06	(4.45)&	8.76E-01	&	2.97E-02	&	3637	 (5.0)	\\
13 &	T{\bf H}	&	65.79	$_{-0.49	}^{+0.41	}$ &	5.13	$_{-1.44	}^{+1.23	}$ &	0.96	$_{-0.09	}^{+0.08	}$ &	1.34E-09	(5.95)&	4.58E-04	&	6.61E-06	&	5490	 (5.0)	\\
14 &	T{\bf H}	&	65.96	$_{-1.02	}^{+1.17	}$ &	12.03$_{-3.71	}^{+2.82	}$ &	0.91	$_{-0.11	}^{+0.09	}$ &	4.79E-07	(4.90)&	6.75E-02	&	1.01E-03	&	4767	 (5.0)	\\
15 &	T{\bf H}	&	68.07	$_{-0.24	}^{+0.25	}$ &	3.99	$_{-0.73	}^{+0.59	}$ &	1.01	$_{-0.07	}^{+0.06	}$ &	0.00E+00	(8.60)&	0.00E+00	&	0.00E+00	&	3714	 (4.6)	\\
16 &	T{\bf H}	&	68.72	$_{-0.38	}^{+0.37	}$ &	4.90	$_{-1.91	}^{+1.26	}$ &	0.82	$_{-0.10	}^{+0.08	}$ &	3.33E-08	(5.40)&	1.19E-02	&	1.72E-04	&	4021	 (5.0)	\\
17 &	{\bf H}	&	62.65	$_{-1.35	}^{+1.40	}$ &	10.42$_{-2.98	}^{+2.39	}$ &	1.06	$_{-0.13	}^{+0.12	}$ &	3.40E-06	(4.50)&	6.81E-01	&	1.64E-02	&	4324	 (4.6)	\\
18 &	{\bf H}	&	58.90	$_{-1.80	}^{+1.67	}$ &	17.62$_{-4.87	}^{+3.91	}$ &	1.35	$_{-0.16	}^{+0.15	}$ &	3.40E-06	(4.50)&	4.91E-01	&	9.71E-03	&	3922	 (5.0)	\\
19 &	T{\bf H}	&	64.75	$_{-0.33	}^{+0.31	}$ &	3.91	$_{-1.11	}^{+0.86	}$ &	0.62	$_{-0.06	}^{+0.05	}$ &	1.82E-09	(5.90)&	8.15E-04	&	1.18E-05	&	5982	 (4.6)	\\
20 &	{\bf T}H	&	65.89	$_{-0.53	}^{+0.51	}$ &	3.87	$_{-2.14	}^{+1.48	}$ &	0.50	$_{-0.09	}^{+0.07	}$ &	4.81E-05	(3.90)&	1.00E+00	&	2.70E-01	&	4349	 (5.0)	\\
21 &	T{\bf H}	&	64.99	$_{-0.24	}^{+0.24	}$ &	3.21	$_{-0.76	}^{+0.64	}$ &	0.66	$_{-0.05	}^{+0.05	}$ &	3.01E-13	(7.20)&	1.64E-07	&	2.37E-09	&	6235	 (5.0)	\\
22 & {\bf T}	&	65.45	$_{-0.95	}^{+0.82	}$ &	2.55	$_{-1.95	}^{+1.75	}$ &	0.79	$_{-0.13	}^{+0.13	}$ &	1.26E-03	(3.02)&	1.00E+00	&	1.00E+00	&	2826	 (3.0)	\\
23 &	T{\bf H}	&	70.19	$_{-0.42	}^{+0.49	}$ &	3.76	$_{-1.80	}^{+1.36	}$ &	1.03	$_{-0.15	}^{+0.13	}$ &	1.33E-05	(4.20)&	9.98E-01	&	8.60E-02	&	3099	 (3.1)	\\
24 &	{\bf H}	&	70.39	$_{-0.48	}^{+0.49	}$ &	2.97	$_{-1.94	}^{+2.10	}$ &	1.17	$_{-0.22	}^{+0.18	}$ &	2.33E-04	(3.50)&	1.00E+00	&	9.69E-01	&	3301	 (3.0)	\\
25 &	{\bf H}	&	68.92	$_{-0.41	}^{+0.57	}$ &	3.66	$_{-2.05	}^{+1.49	}$ &	0.91	$_{-0.15	}^{+0.13	}$ &	1.08E-04	(3.70)&	1.00E+00	&	7.75E-01	&	3799	 (4.0)	\\
26 &	T{\bf H}	&	68.89	$_{-0.67	}^{+0.56	}$ &	5.41	$_{-1.68	}^{+1.31	}$ &	1.39	$_{-0.17	}^{+0.14	}$ &	3.33E-08	(5.40)&	1.07E-02	&	1.56E-04	&	2814	 (3.0)	\\
27 &	T{\bf H}	&	68.82	$_{-0.39	}^{+0.39	}$ &	3.82	$_{-1.38	}^{+1.17	}$ &	1.66	$_{-0.19	}^{+0.17	}$ &	2.87E-07	(5.00)&	1.23E-01	&	1.90E-03	&	3626	 (3.0)	\\
28 &	{\bf H}	&	70.26	$_{-0.45	}^{+0.45	}$ &	2.98	$_{-1.41	}^{+1.00	}$ &	1.32	$_{-0.21	}^{+0.18	}$ &	1.08E-04	(3.70)&	1.00E+00	&	8.39E-01	&	3536	 (3.0)	\\
29 &	{\bf H}	&	68.57	$_{-1.40	}^{+1.40	}$ &	7.04	$_{-4.34	}^{+2.87	}$ &	1.40	$_{-0.22	}^{+0.22	}$ &	7.36E-04	(3.18)&	1.00E+00	&	1.00E+00	&	3481	 (3.0)	\\
30 &	{\bf H}	&	69.58	$_{-0.54	}^{+0.47	}$ &	3.01	$_{-1.97	}^{+1.42	}$ &	1.29	$_{-0.26	}^{+0.19	}$ &	3.37E-04	(3.40)&	1.00E+00	&	9.41E-01	&	3678	 (3.0)	\\
31 &	T{\bf H}	&	69.31	$_{-0.32	}^{+0.34	}$ &	2.95	$_{-1.09	}^{+0.86	}$ &	1.44	$_{-0.16	}^{+0.15	}$ &	4.79E-07	(4.90)&	2.48E-01	&	4.10E-03	&	3656	 (3.0)	\\
32 &	T{\bf H}	&	69.61	$_{-0.76	}^{+1.08	}$ &	6.60	$_{-4.34	}^{+2.31	}$ &	1.04	$_{-0.21	}^{+0.15	}$ &	2.33E-04	(3.50)&	1.00E+00	&	5.90E-01	&	4235	 (3.0)	\\
33 &	{\bf T}	&	134.10	$_{-1.80	}^{+1.88	}$ &	6.81	$_{-2.96	}^{+2.20	}$ &	2.00	$_{-0.33	}^{+0.30	}$ &	3.37E-04	(3.40)&	1.00E+00	&	9.18E-01	&	1944	 (2.0)	\\
34 &	T{\bf H}	&	67.99	$_{-1.04	}^{+1.10	}$ &	5.28	$_{-3.78	}^{+1.97	}$ &	1.63	$_{-0.21	}^{+0.21	}$ &	5.22E-05	(3.88)&	1.00E+00	&	2.21E-01	&	2455 (4.0)	\\
35 &	{\bf T}H	&	142.98	$_{-3.15	}^{+3.81	}$ &	18.71$_{-9.32	}^{+6.40	}$ &	1.90	$_{-0.32	}^{+0.27	}$ &	2.33E-04	(3.50)&	1.00E+00	&	2.70E-01	&	2544	 (2.0)	\\
36 &	{\bf T}	&	63.35	$_{-0.99	}^{+0.89	}$ &	4.08	$_{-1.85	}^{+1.72	}$ &	0.81	$_{-0.15	}^{+0.13	}$ &	1.35E-03	(3.00)&	1.00E+00	&	1.00E+00	&	3051	 (3.0)	\\
37 &	T{\bf H}	&	66.33	$_{-0.15	}^{+0.16	}$ &	1.72	$_{-0.57	}^{+0.39	}$ &	0.82	$_{-0.08	}^{+0.08	}$ &	3.33E-08	(5.40)&	3.35E-02	&	4.91E-04	&	5834	 (3.0)	\\
38 &	T{\bf H}	&	68.31	$_{-0.09	}^{+0.09	}$ &	2.61	$_{-0.28	}^{+0.25	}$ &	2.64	$_{-0.09	}^{+0.09	}$ &	0.00E+00	(15.00)&	0.00E+00	&	0.00E+00	&	3245	 (3.0)	\\
39 &	T{\bf H}	&	67.77	$_{-0.10	}^{+0.10	}$ &	2.99	$_{-0.23	}^{+0.21	}$ &	3.13	$_{-0.09	}^{+0.09	}$ &	0.00E+00	(18.50)&	0.00E+00	&	0.00E+00	&	3341	 (3.0)	\\
40 &	T{\bf H}	&	67.82	$_{-0.14	}^{+0.15	}$ &	2.94	$_{-0.32	}^{+0.32	}$ &	3.13	$_{-0.12	}^{+0.11	}$ &	0.00E+00	(13.80)&	0.00E+00	&	0.00E+00	&	3210	 (3.0)	\\
41 &	T{\bf H}	&	65.43	$_{-0.60	}^{+0.78	}$ &	5.14	$_{-2.25	}^{+1.66	}$ &	0.76	$_{-0.11	}^{+0.09	}$ &	2.07E-05	(4.10)&	9.99E-01	&	9.67E-02	&	5584	 (2.8)	\\
42 &	{\bf T}H	&	66.05	$_{-0.46	}^{+0.45	}$ &	2.92	$_{-1.56	}^{+1.28	}$ &	0.73	$_{-0.12	}^{+0.00	}$ &	2.33E-04	(3.50)&	1.00E+00	&	8.67E-01	&	1937	 (3.0)	\\
43 &	T{\bf H}	&	65.91	$_{-0.71	}^{+0.62	}$ &	6.24	$_{-3.13	}^{+2.04	}$ &	0.98	$_{-0.14	}^{+0.12	}$ &	1.33E-05	(4.20)&	9.76E-01	&	5.27E-02	&	6289	 (3.0)	\\
44 &	T{\bf T}	&	67.74	$_{-1.03	}^{+1.01	}$ &	3.90	$_{-3.23	}^{+2.06	}$ &	0.71	$_{-0.11	}^{+0.11	}$ &	6.41E-04	(3.22)&	1.00E+01	&	1.00E+01	&	4992 (2.0)	\\
45 &	{\bf T}	&	67.17	$_{-0.35	}^{+0.49	}$ &	2.26	$_{-1.44	}^{+0.92	}$ &	1.06	$_{-0.19	}^{+0.16	}$ &	3.37E-04	(3.40)&	1.00E+00	&	9.99E-01	&	2516	 (2.0)	\\
46 &	{\bf T}H	&	67.52	$_{-0.39	}^{+0.36	}$ &	2.38	$_{-1.15	}^{+0.79	}$ &	1.04	$_{-0.17	}^{+0.14	}$ &	7.23E-05	(3.80)&	1.00E+00	&	5.37E-01	&	2570	 (2.0)	\\
47 &	{\bf T}H	&	64.52	$_{-0.27	}^{+0.23	}$ &	3.15	$_{-1.35	}^{+0.85	}$ &	0.49	$_{-0.06	}^{+0.05	}$ &	2.87E-07	(5.00)&	1.47E-01	&	2.30E-03	&	6318	 (3.3)	\\
48 &	{\bf H}	&	71.30	$_{-0.62	}^{+0.66	}$ &	4.26	$_{-2.80	}^{+1.28	}$ &	1.73	$_{-0.36	}^{+0.22	}$ &	5.41E-06	(3.92)&	9.88E-01	&	5.47E-02	&	3523	 (2.0)	\\
49 &	{\bf T}H	&	67.21	$_{-0.42	}^{+0.35	}$ &	3.20	$_{-1.65	}^{+1.40	}$ &	1.30	$_{-0.18	}^{+0.15	}$ &	5.41E-06	(4.40)&	9.48E-01	&	4.19E-02	&	2144	 (2.0)	\\
50 &	{\bf T}	&	66.54	$_{-1.08	}^{+1.07	}$ &	8.40	$_{-5.12	}^{+3.02	}$ &	1.75	$_{-0.35	}^{+0.26	}$ &	2.33E-04	(3.50)&	1.00E+00	&	7.54E-01	&	1658	 (2.0)	\\
51 &	{\bf T}	&	68.89	$_{-0.61	}^{+0.62	}$ &	2.60	$_{-2.85	}^{+8.03	}$ &	1.27	$_{-0.33	}^{+0.22	}$ &	1.35E-03	(3.00)&	1.00E+00	&	1.00E+00	&	2058	 (2.0)	\\
 
\hline
\end{tabular}
\label{tab:results}
\end{minipage}
\end{table*}

\begin{figure}
\begin{center}
\includegraphics[width=8.5cm]{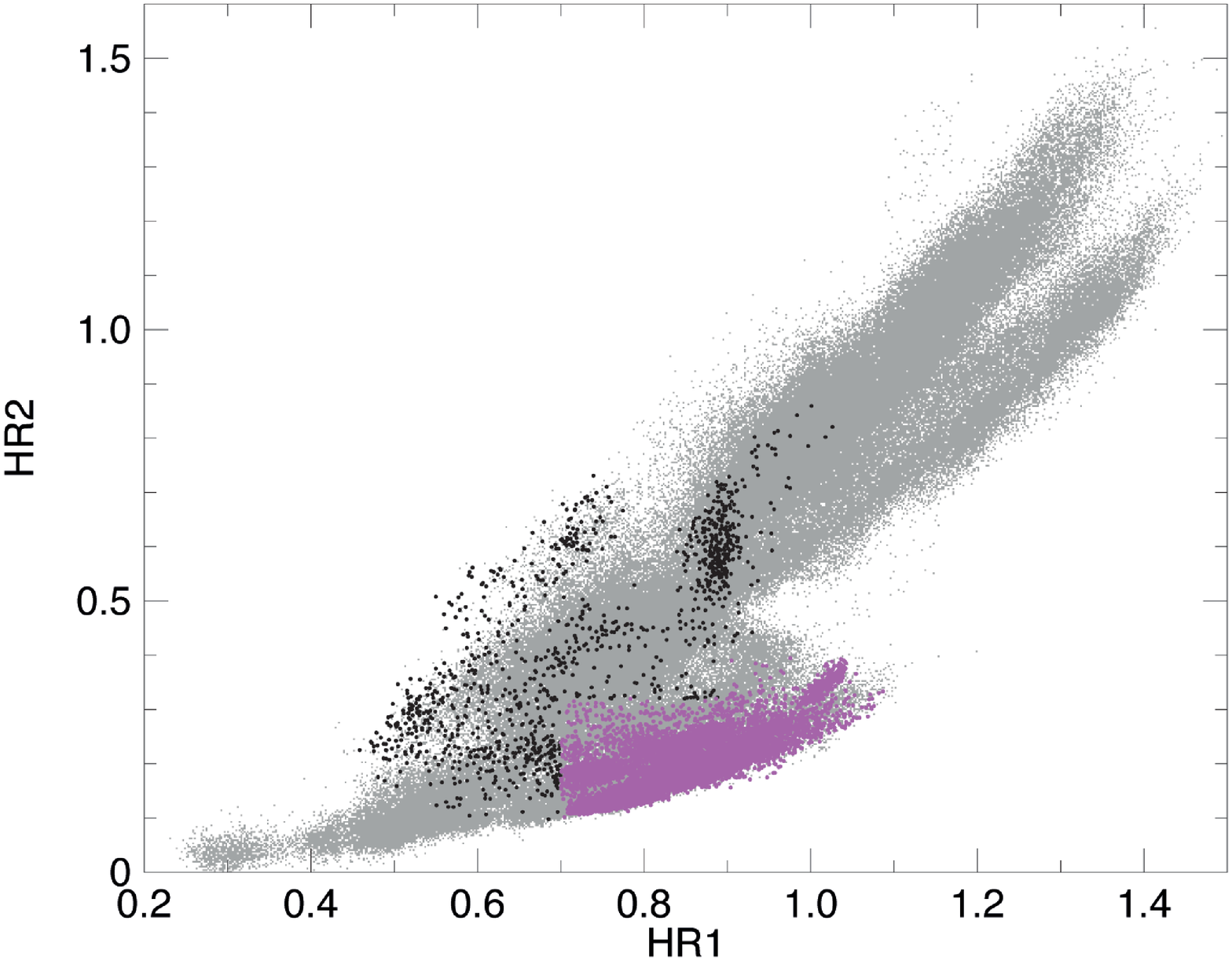}\\
\includegraphics[width=8.5cm]{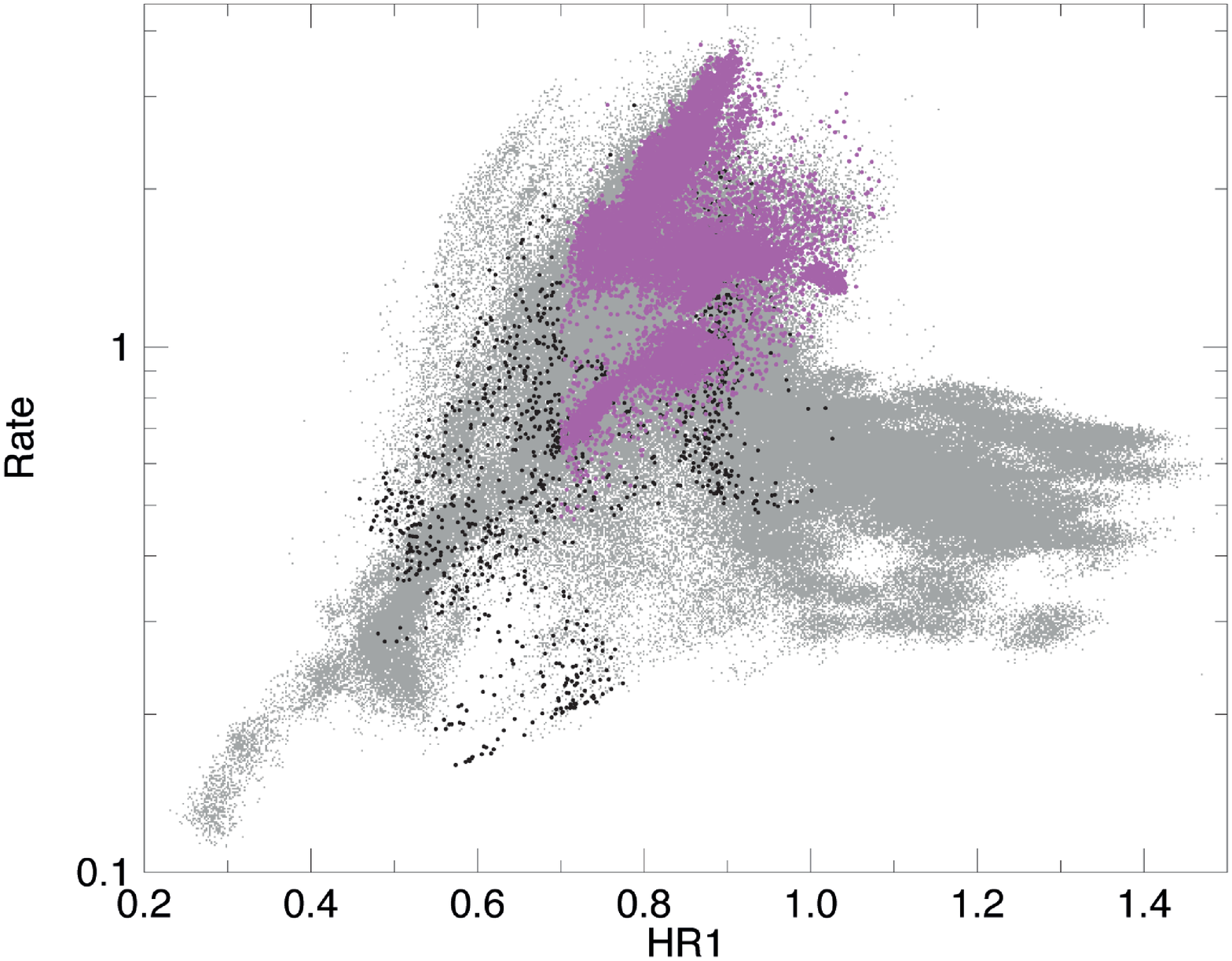}\\
\end{center}
\caption{Top: Colour-colour diagram for all RXTE/PCA observations of GRS 1915+105. Each point is accumulated for 16 seconds. The magenta \revised{and black} circles mark all segments included in the observations where a HFQPO was detected. Different colours mark point in different areas (see text). Bottom: corresponding hardness-intensity diagram.
}
\label{fig:allcolors}
\end{figure}

\subsection{Colour--colour and hardness--intensity diagrams}

We use the 16-s time-resolution Standard 2 mode data to calculate X-ray colours.  
For each of the five PCA detectors (PCUs) we extracted count rates every 16 seconds in the 2.0--6.0 keV, 6.0--16.0 keV, 16.0--20.0 keV and 2.0--20 keV bands.
To obtain the count rates in the exact energy ranges, we interpolated linearly between PCU channels.
Each light curve was corrected for dead-time effects following the methods suggested by the RXTE team. 
We also remove instrumental drop-outs and subtracted the background contribution in each band using the standard bright source background model for the PCA, version 3.8\footnote{PCA Digest at http://heasarc.gsfc.nasa.gov/ for details of the model}.
We defined the soft colour (HR1) as the count rate in the 6.0--16.0 keV band divided by the rate in the 2.0--6.0 keV band and the hard colour (HR2) as the ratio of the count rates in the 16.0--20.0 keV rate divided by the 2.0--6.0 keV rate.
We used the count rates in the 2.0--20 keV band as a measurement of the source intensity.
In the 16 years of activity, the RXTE gain changed 5 times with each new high voltage setting of the PCUs (Jahoda et al. 2006).
In order to correct for these gain changes as well as the differences in effective area between the PCUs, we used the method introduced by Kuulkers et al. (1994): for each PCU we calculate, in the same manner as for GRS~1915+105, the colours of the Crab, which can be supposed to be constant. 
We then average the 16s Crab colours and intensity for each PCU for each day. For each PCU we divide the 16s colour and intensity values obtained for GRS~1915+105 by the corresponding average Crab values that are closest in time but in the same RXTE gain epoch. Then, we average the colours and intensity over all active PCUs to obtain colours and intensities every 16 seconds.

The colour-colour diagram (CCD) and hardness-intensity diagram (HID) obtained with the 16s points are shown in Fig. \ref{fig:allcolors}. The colour dots marks the 16s segments belonging to observations where we detected a HFQPO.
 The colours are assigned in the following way: 
\begin{itemize}
\item {\it magenta}: (HR$_1$ $>$ 0.7 \& HR$_2$ $<$ 0.32) \revised{or} (HR$_1$ $>$ 0.9 \& HR$_2$ $<$ 0.4)
\item {\it black}: remaining points.
\end{itemize}

\section{Results}

\subsection{Where HFQPOs are detected}

\begin{figure}
\begin{center}
\includegraphics[width=8.5cm]{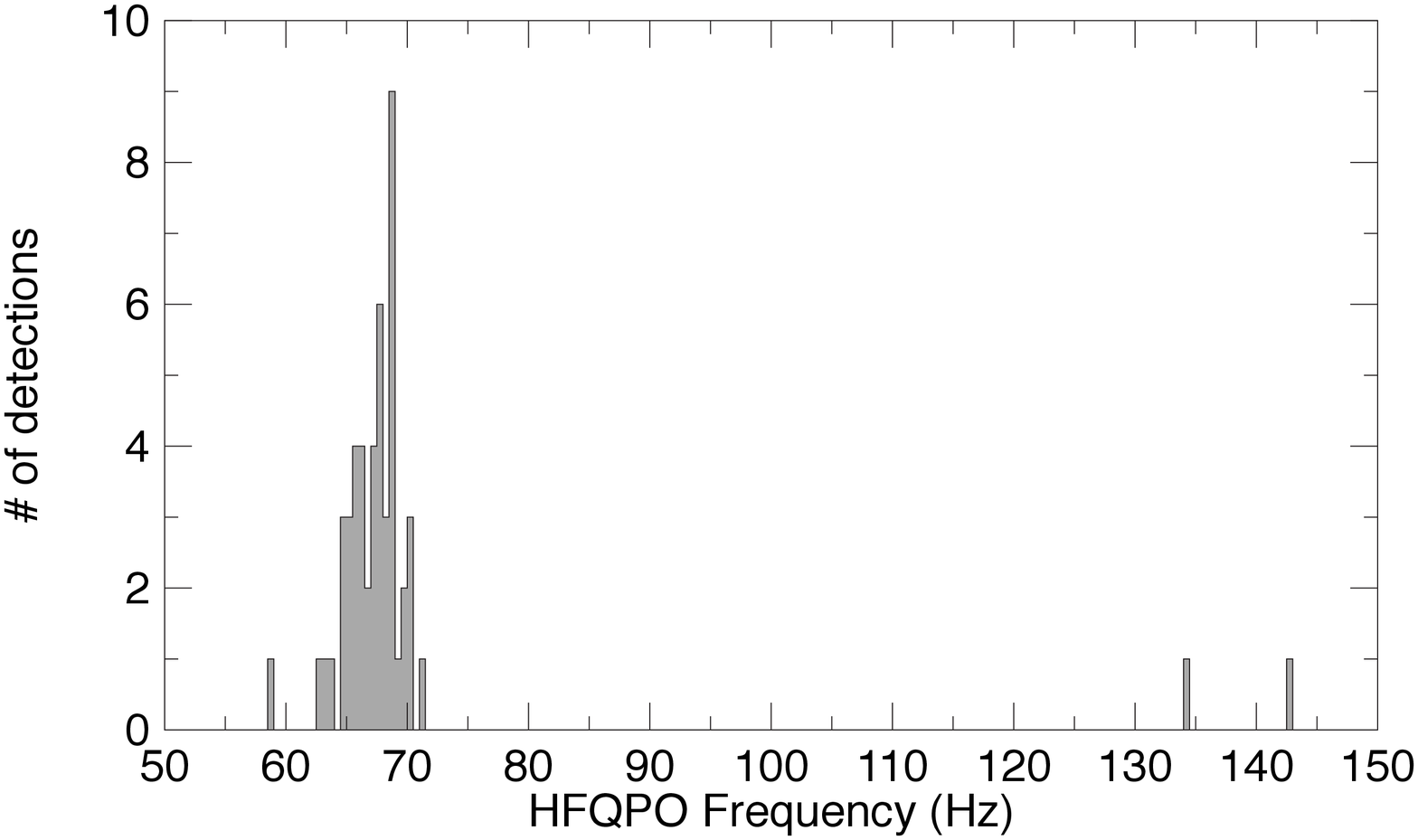}\\
\includegraphics[width=8.5cm]{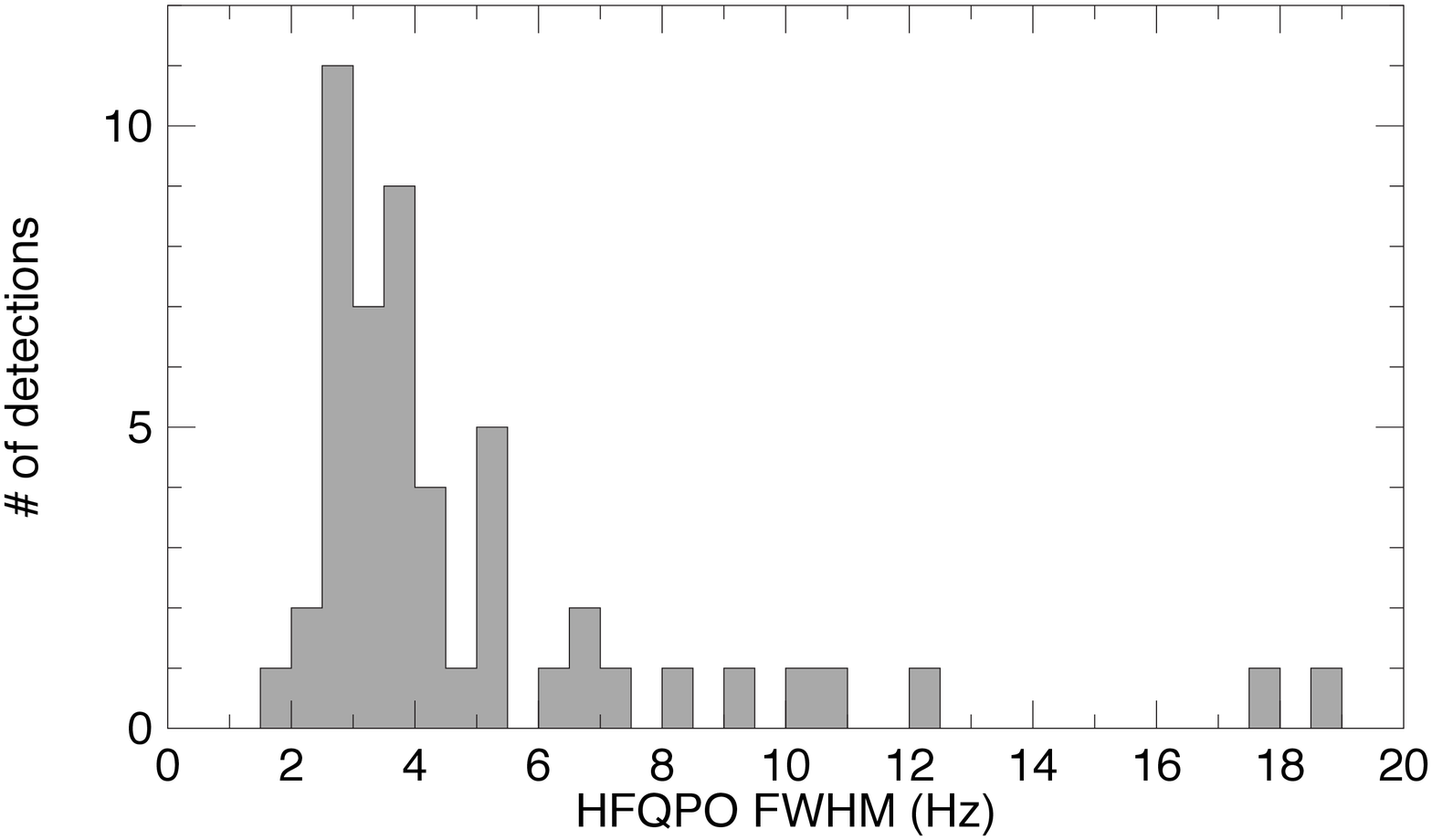}\\
\includegraphics[width=8.5cm]{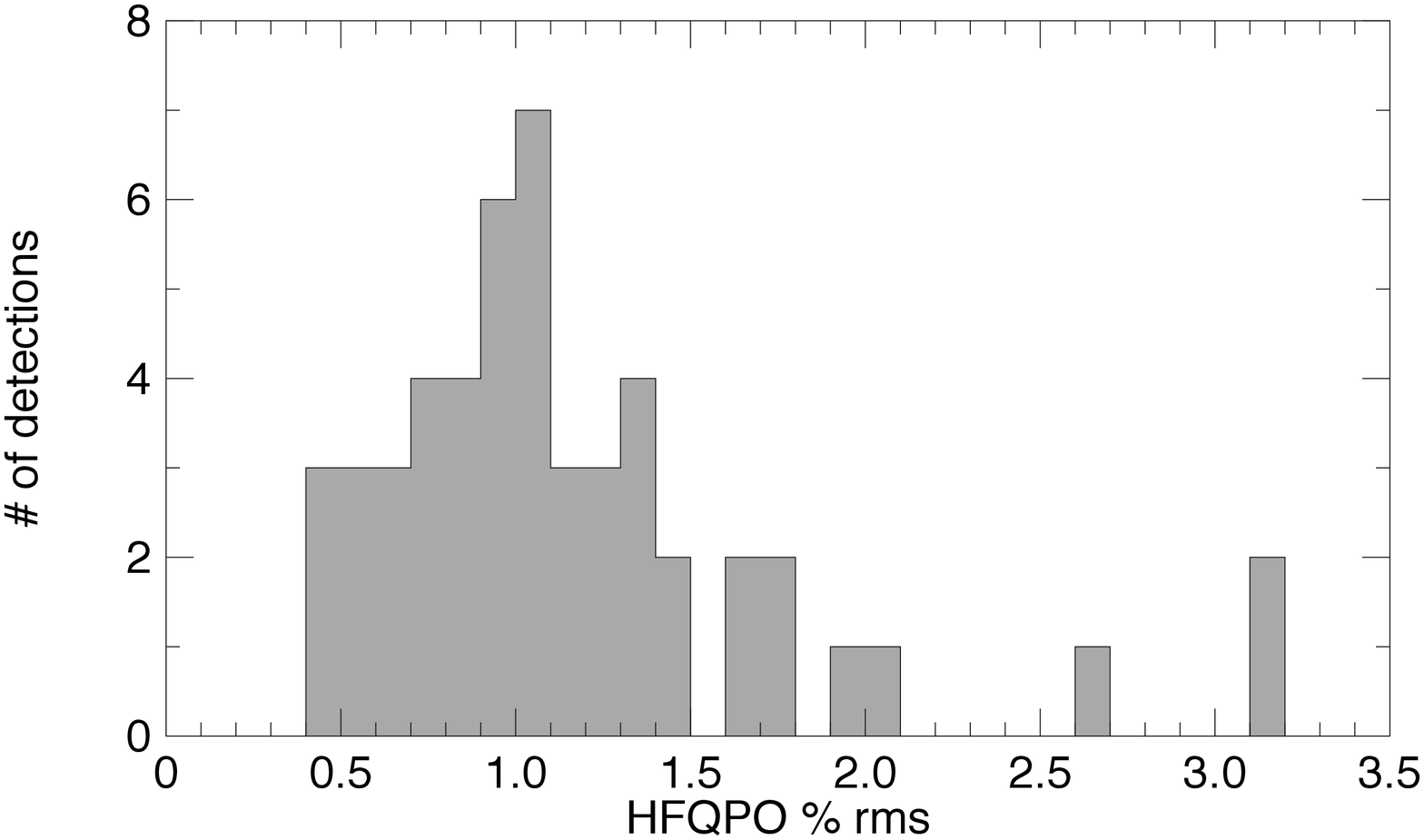}\\
\end{center}
\caption{Distributions of best-fit parameters for all HFQPOs detected (see Tab. \ref{tab:results}). From top to bottom: centroid frequency, FWHM and fractional percentage rms.
}
\label{fig:histograms}
\end{figure}

\begin{figure}
\begin{center}
\includegraphics[width=8.5cm]{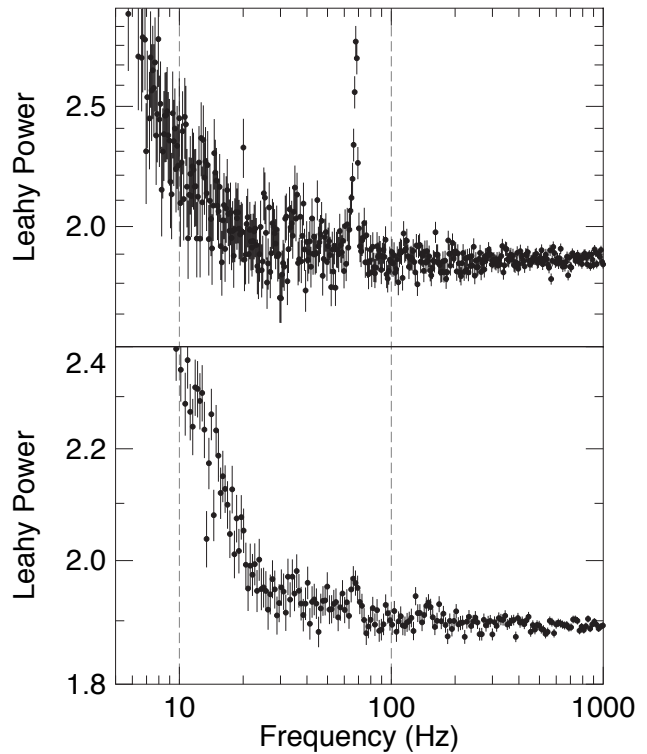}
\end{center}
\caption{Plot of our most significant (top panel, \revised{Obs. \#39)} and \revised{one of the} least significant (bottom panel, Obs. \#1) HFQPO detections.
}
\label{fig:bestworst}
\end{figure}

Figure \ref{fig:histograms} shows the distribution of the parameters of our detected peaks. The centroid frequencies are concentrated around 67 Hz, with \revised{48 our of 51} detections between 63.5 and 71.3 Hz, with an average of 67.3$\pm$2.0 Hz. The FWHM peaks between 3 and 4 Hz, with an average of 4.4$\pm$2.4 Hz. The fractional rms, which is of course affected by our sensitivity in its lowest values, peaks at 1 \%.
In Fig. \ref{fig:bestworst} we show two examples of detections: that with the highest single-trial significance (Obs. 34, $n_\sigma$=18.5) and one of the 3$\sigma$ ones (Obs. 1, $n_\sigma$=3).

Our \revised{51} detections listed in Tab. \ref{tab:results} come from observations from seven variability classes: $\kappa$,$\gamma$,$\mu$,$\delta$,$\omega$,$\rho$,$\nu$ (see Belloni et al. 2000; Klein-Wolt et al. 2002). 
For many of these, the average rates quoted in Tab. \ref{tab:results} are only indicative, as there are large rate variations during the observation.
Noticeably, none of the detections corresponds to observations in class $\chi$, which is the one showing consistently a type-C QPO. This is in agreement with the results of Belloni, Sanna \& M\' endez (2012), who found no HFQPO and type-C detection in other sources.
Belloni et al. (2000) have shown that the $>$1s variability of GRS 1915+105 can be reduced to the transition between three basic spectral states, called A, B and C. Only two variability classes do not involve state transitions: $\phi$ and $\chi$, none of which appear in our sample. 

In order to establish whether the presence of the HFQPO is related to a specific state, we extracted PDS in the total band from selections made on the HID and CCD in Fig. \ref{fig:allcolors}.
All 16s segments from the observations where we detect a HFQPO  are shown in colour in Fig. \ref{fig:allcolors}. We identified \revised{two} regions (for their precise definition see above): the soft points (magenta), which occupy a dense region of the CCD, \revised{and the remaining points (black).}

\begin{figure*}
\begin{center}
\begin{tabular}{cc}
\includegraphics[width=8.5cm]{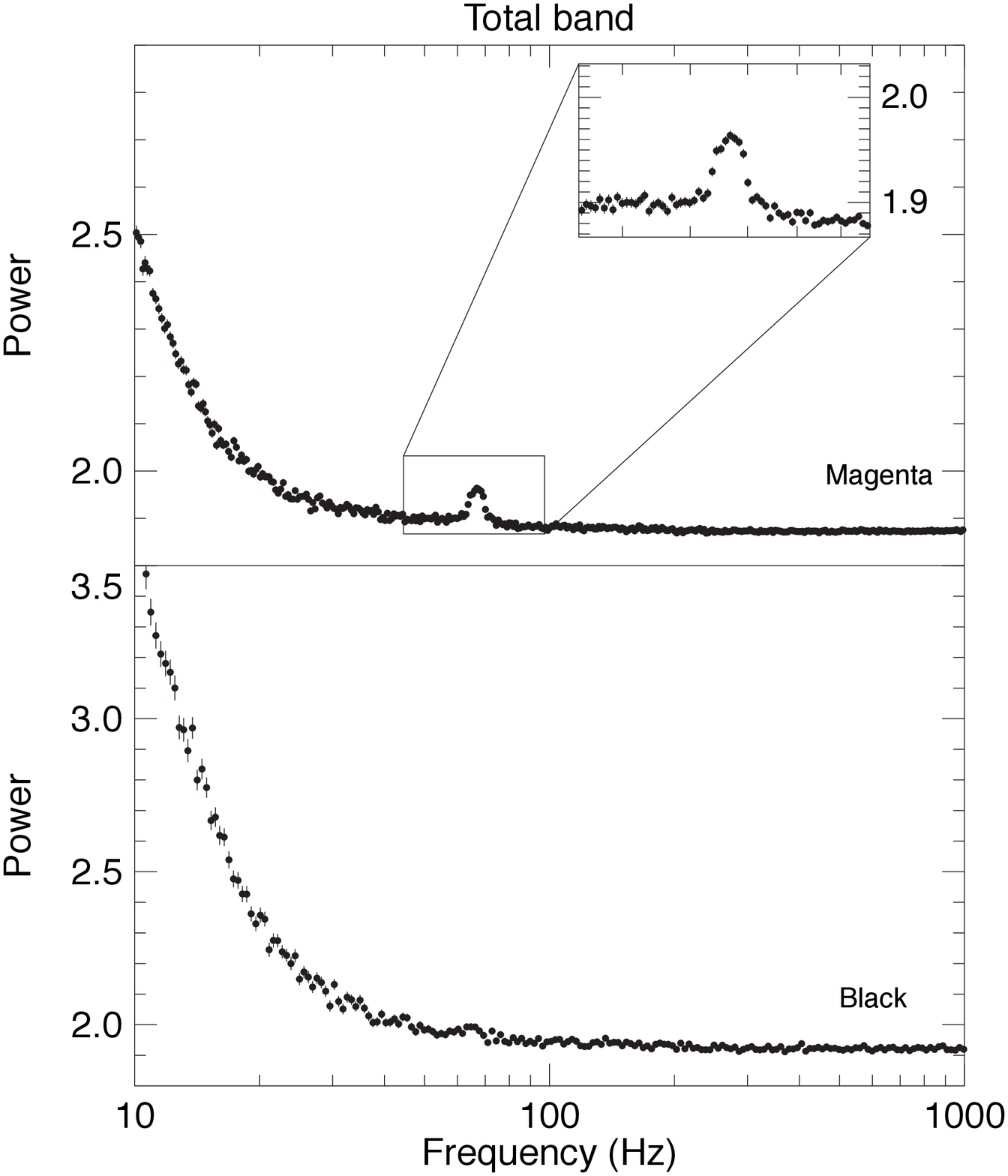} &
\includegraphics[width=8.5cm]{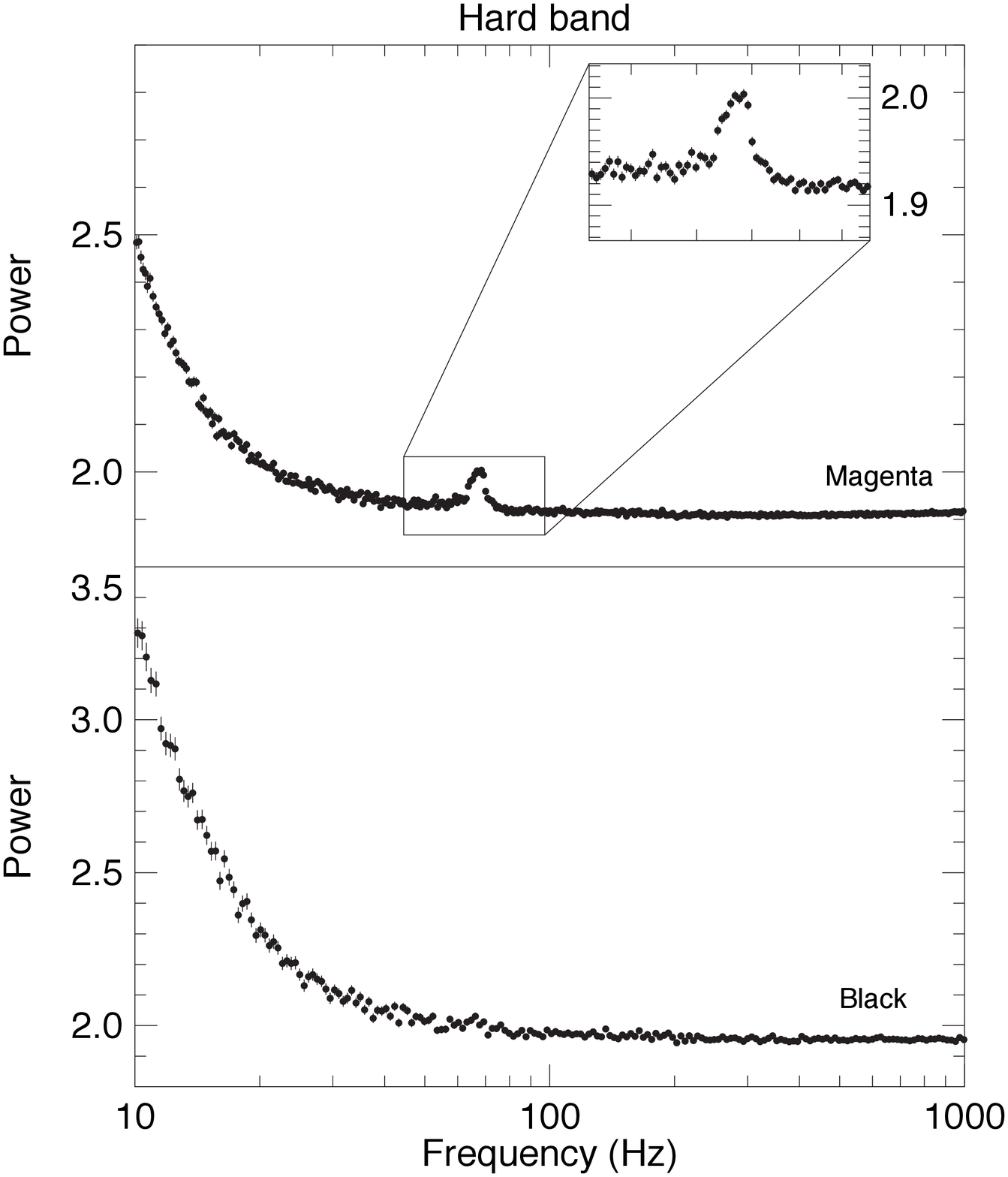} \\
\end{tabular}
\end{center}
\caption{PDS averaged over the colour regions in Fig. \ref{fig:allcolors} \revised{(corresponding to the observations where a HFQPO is detected)} \revised{for the total and hard band (left and right, respectively).} The insets in the top panels shows a zoom around the QPO area. No QPOs are detected in the bottom panels. \revised{Error bars in the top panels are smaller than the symbols.}
}
\label{fig:color_detections}
\end{figure*}

We extracted the average PDS from the \revised{two} regions \revised{from both the total and hard bands}, shown in Fig. \ref{fig:color_detections}. It is clear that the HFQPO is detected only in the magenta region, which corresponds to state B (the CCD in Fig. \ref{fig:allcolors} is comparable with that in Belloni et al. 2000, with axes flipped). 
The black points cover both a softer region of state B and part of state C. No state A point is either black or magenta.
The magenta selection yields an averaged QPO centered \revised{at 66.9$\pm$0.1 Hz, with a FWHM of 5.5$\pm$0.3 Hz and a fractional rms of 0.69$\pm$0.02\% for the full band (significance 24.4$\sigma$) and centered at 67.1$\pm$0.1 Hz, with a FWHM of 5.9$\pm$0.3 Hz and a fractional rms of 0.93$\pm$0.02\% for the hard band (significance 26.3$\sigma$).}
The \revised{black selection, assuming the same shape for the QPO, yields 3$\sigma$ upper limits of 0.8\% (total band) and 1.0\% (hard band).}

\begin{figure}
\begin{center}
\includegraphics[width=8.5cm]{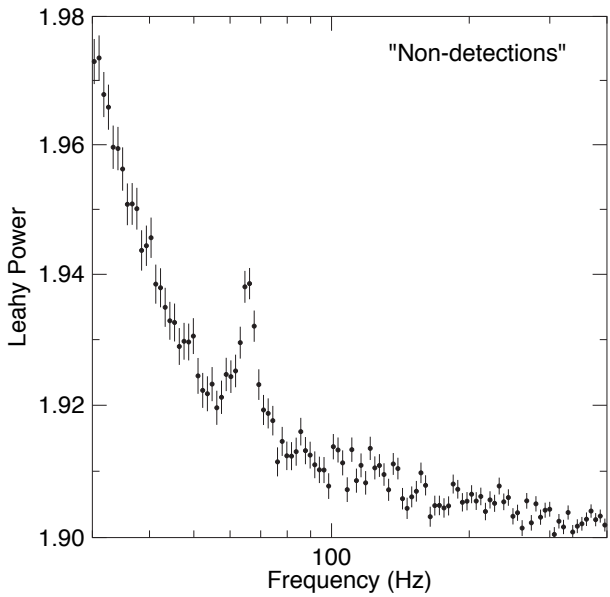}
\end{center}
\caption{PDS obtained from the average of all 16s points in the magenta region, but belonging to observations where we do not detect a HFQPO in the single full observations. 
}
\label{fig:non-detection}
\end{figure}

Having identified the region in the CCD where HFQPOs are found, we accumulated a PDS from all 16s points in that region from observations where we did not detect a HFQPO in single observations. This selection resulted in a positive 7.9$\sigma$ detection (see Fig. \ref{fig:non-detection}), indicating that state B is indeed associated to the presence of HFQPO. The centroid is 65.8$\pm$0.3 Hz, the FWHM 4.3$\pm$0.6 Hz and the integrated fractional rms 0.33$\pm$0.02\%, lower by roughly a factor of three than that from the magenta points. \revised{It corresponds to a significance of 8.3$\sigma$.} 
\revised{Not all PDS averaged need to contain an identical signal and therefore this value had to be taken as an average value. If there are PDS with a weaker signal, this value must be seen as an upper limit.}
Notice that also the centroid frequency is significantly lower, although it is difficult to say anything more as these are averages over a large number of observations.
The same result was obtained by Altamirano \& Belloni (2011) for IGR J10791-3624. The HQFPO in that source was found only in a subset of bright observations; selecting all remaining bright observations did reveal a similar QPO peak.

\revised{In order to check for a possible correlation between QPO frequency and rate, we averaged the Crab-corrected rates from the magenta points observation by observation, obtaining a clean count rate for each observation, limited to the intervals where the HFQPO is detected. A linear regression finds a correlation factor of -0.29, indicating the absence of a statistically significant correlation.}

\subsection{Rms spectrum of the HFQPO}

\begin{figure}
\begin{center}
\includegraphics[width=8.5cm]{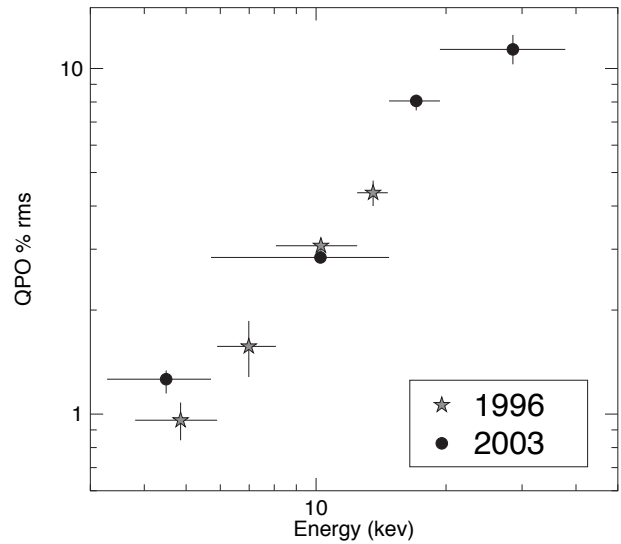}
\end{center}
\caption{Rms spectra of the HFQPO for observation \#5 (black dots, 1996, see also Morgan et al. 1997) and for the combined observations \revised{\#38,\#39,\#40} (gray stars, 2003).
}
\label{fig:rms}
\end{figure}

In order to extract the rms spectrum of the QPO we selected two groups of observations. The first is single observation \#5 (from 1996), the same analyzed by Morgan et al. (1997). The second is the combination of observations \revised{\#38,\#39,\#40} (from 2003), which are from the same day and all show a very strong HFQPO peak at a very similar frequency. The resulting rms spectra are shown in Fig. \ref{fig:rms}. The rms vs. energy dependence is very similar and with the 2003 observations, \revised{but we are able to extend it} to 40 keV, showing that it continues to increase up to at least those energies, with an rms larger than 10 \%.

\subsection{The 41 Hz QPO}

Strohmayer (2001a) reported the discovery of a second QPO peak around 41 Hz for a set of five observations in 1997. We do not see this peak in single observations, although in each of them we detect a $\sim$67 Hz HFQPO. However, averaging those five observations we reproduce the 41 Hz detection. These observations cover a subset of colours of the magenta area. Selecting all other points in this subset does not reveal the oscillation. The conclusion is that the 41 Hz feature is a \revised{much more} transient phenomenon, unlike the 67 Hz one, which appears \revised{much more frequently}.

\section{Discussion}

We have performed a full analysis of 1807 RXTE/PCA observations of GRS 1915+105 with the aim of obtaining a homogeneous set of detections of its HFQPOs. Our procedure returned \revised{51} observations with a single-trial detection significance larger than 3$\sigma$. All but three of these detections correspond to centroid frequencies between 63.5 and 71.3 Hz, one is at a lower frequency (59 Hz) and two are at higher frequencies (127 and 143 Hz). The latter are broadly consistent with being a factor of two higher in frequency than the $\sim$67 Hz main group.
The \revised{48} detections are grouped around a centroid frequency of $\sim$67 Hz, a FWHM of $\sim$3 Hz and have a total fractional rms of 0.5-1.5\%. 67 Hz appears to be a very special frequency associated to this particular source.
An analysis made selecting specific intensities and X-ray colours indicate that the HFQPO is associated only to state B of GRS 1915+105 (see Belloni et al. 2000).
\revised{Three separate states have been defined for the variability of GRS 1915+105 (Belloni et al. 2000). State C is equivalent to the Hard Intermediate State of more conventional black-hole transients, where the spectrum is composed by a thermal disk component plus a relatively strong hard tail, which can dominate the flux (see Belloni 2010; Belloni, Motta \& Mu\~noz-Darias 2012). States A and B correspond to an energy spectrum strongly dominated by the thermal disk component and have a PDS almost featureless (Reig, Belloni \& van der Klis 2003) and can be compared to the ``anomalous'' state of black-hole binaries (Belloni 2010). Here  the energy spectrum is dominated by a bright thermal component and the inner radius of the accretion disk is close to its lower value (Belloni et al. 1997a,b; 2000). We obtain no detection during state C}, when strong type-C QPOs are observed (see Markwardt, Swank \& Taam 1999; Fender \& Belloni 2004 and references therein). However, the rms spectrum of the HFQPO is very hard and increases up to 40 keV without any sign of flattening (see also Morgan et al. 1997).

It is remarkable that the frequency we find is very close to that discovered from the only known X-ray source which displays properties similar to those of GRS 1915+105, IGR J17091-3624 (Altamirano \& Belloni 2012). The frequency recovered when analysing all magenta points in GRS 1915+105 is 67.1$\pm$0.1 Hz, while the detection in IGR J17091-3624 is at 66.5$\pm$0.5 Hz, compatible within 1$\sigma$. Altamirano et al. (2012) suggested that IGR J17091-3624 has a lower mass than GRS 1915+105, but even assuming a similar mass the detection of precisely the same frequency is puzzling.

RXTE observed dozens of black-hole transients in its sixteen years of operation. Overall, the number of detections of HFQPOs from these systems was very limited (Belloni, Sanna \& M\'endez 2012). GRS 1915+105 alone yielded more than 40 detections, which seems to deviate from the pattern. However, it was shown that the detectability of HFQPOs strongly depends on source state. All detections in other transients were associated to relatively rare states at high luminosity (the Soft-Intermediate and Anomalous states, see Belloni 2010). GRS 1915+105 does not follow the same pattern as conventional transients, but it is extremely bright and it was shown that its properties can be associated to the same basic states (see e.g. Reig et al. 2003; Soleri et al. 2008; Belloni 2010). Although it is difficult to make a quantitative comparison, it is clear that GRS 1915+105 spends a much higher percentage of time in HFQPO-related states and therefore its properties might be compatible with those of the others. 

The HFQPO frequency cannot be the Keplerian frequency at the Innermost Stable Orbit (ISCO) around the black hole in GRS 1915+105. For a dynamical mass of 14.0$\pm$4.4 M$_\odot$ (Harlaftis \& Greiner 2004), the lowest possible Keplerian frequency at the ISCO, obtained in the case of M=18.8 M$_\odot$ and no spin, is 110 Hz. In order to reach 67 Hz, the black hole mass should be as high as 30 M$_\odot$. However, theoretical models proposed for the HFQPOs do not \revised{always} associate the frequency to the ISCO.
The relativistic resonance model (Kluzniak \& Abramowicz 2002) associates the QPO frequencies to relativistic time scales at a specific radius, where these frequencies are in resonance, resulting in special frequency ratios. We found only single peaks in our analysis, but the fact that the variations in this frequency over 16 years of RXTE life was very small is qualitatively compatible with that idea. From Fig. \ref{fig:histograms} the QPO frequency is distributed over a narrow range of frequencies, but even limiting ourselves to the main peak in the distribution it varies $\pm$4 Hz, i.e. 6\% These variations must be evaluated in the framework of the model.
The relativistic precession model (see Stella, Vietri \& Morsink 1999 and references therein) makes predictions about the frequencies of both high-frequency peaks. Having only a single peak does not allow any testing of this model.
The same applies to other models (disko-seismic: Nowak \& Wagoner 1991; Accretion-Ejection Instability: Tagger \& Varni\`ere 2006 ; Inner-Torus Oscillation: Rezzolla et al. 2003).

\revised{
There is a detection of a high-frequency feature at a higher frequency in GRS 1915+105, 170 Hz (Remillard et al. 2002b, Belloni et al. 2006), although with a low quality factor. This feature was found by adding observations belonging to variability class $\theta$ and selecting only the hard intervals, corresponding to state C. This is very different from all other detections, in fact class $\theta$ does not even feature state-B intervals, but it is made by oscillations between states A and C (Belloni et al. 2000). This also means that the 170 Hz feature is simultaneous with a type-C QPO, always present in state C.
170 Hz is high enough to be a Keplerian frequency at the ISCO around the black hole in GRS 1915+105, but it is clearly not the main feature observed in this source.
}

Belloni, Sanna \& M\'endez (2012) compared the distribution of HFQPO detected in black-hole transients with that of kHz QPOs in neutron-star binaries, using as comparison the large number of detections obtained for 4U 1636-63 (Sanna et al. 2012). \revised{From the small number of detections, we cannot exclude that the two samples come from a similar distribution}. In our case, the number of detections is larger and concentrated in a narrow range around 67 Hz. Comparing the histogram in Fig. \ref{fig:histograms} with their Fig. 5, the two distributions appear to be markedly different, although a number of selection effects could be at work.

In 2013, the indian X-ray astronomy satellite ASTROSAT will be in operation and will pick up RXTE's legacy on fast timing analysis (Agrawal 2006). In particular, the higher effective area above 20 keV of the Large Area Xenon Proportional Counter (LAXPC) will be ideal for HFQPOs (see Fig. \ref{fig:rms}). GRS 1915+105 in unpredictable in its behaviour, but hardness/intensity selection according to our data will allow to limit the analysis to the most promising intervals.

In the more distant future, if selected by ESA the LOFT satellite will allow us to sample much fainter signals and will most likely lead to many more detections. Depending on the model, different frequencies are expected. However, it is clear that the strongest signal in GRS 1915+105 is around 67 Hz, which appears to be a special value for this source. 
Overall, the results of the complete analysis of HFQPOs in the BHTs observed by RXTE show that these signals are detectd only during specific source states. An observation strategy aimed at these states will allow to maximise detections.

\section*{Acknowledgements}

The research leading to these results has received funding from the European CommunityÕs Seventh Framework Programme (FP7/2007-2013) under grant agreement number ITN 215212 ÔBlack Hole UniverseÕ. 
\revised{This work was partially done during an extended stay of TMB at The University of Southampton, funded by a Leverhulme Trust Visiting Professorship.}

\end{document}